\documentclass[prl,twocolumn,superscriptaddress,showpacs,amsmath,amssymb]{revtex4}
\usepackage[dvipdfm]{graphicx}

\def\x{\mathbf{x}}

\begin{document}
\title{Nodal Domain Statistics for Quantum Maps, Percolation and SLE}
\author{J.P.\ Keating}%
\author{J.\ Marklof}%
\author{I.G.\ Williams}
\affiliation{School of Mathematics, University of Bristol,
Bristol BS8 1TW, UK.}%
\begin{abstract}
We develop a percolation model for nodal domains in the
eigenvectors of quantum chaotic torus maps.  Our model follows
directly from the assumption that the quantum maps are described
by random matrix theory.  Its accuracy in predicting statistical
properties of the nodal domains is demonstrated by numerical
computations for perturbed cat maps and supports the use of
percolation theory to describe the wave functions of general
hamiltonian systems, where the validity of the underlying
assumptions is much less clear.  We also demonstrate that the
nodal domains of the perturbed cat maps obey the Cardy crossing
formula and find evidence that the boundaries of the nodal domains
are described by SLE with $\kappa$ close to the expected value of
6, suggesting that quantum chaotic wave functions may exhibit
conformal invariance in the semiclassical limit.
\end{abstract}
 \maketitle

One of the central problems in the field of quantum chaos is to
understand the morphology of quantum eigenfunctions in classically
chaotic systems.  In time-reversal-symmetric systems one can
always find a basis in which these eigenfunctions are real.  They
can thus be divided into {\it nodal domains} -- connected regions
of the same sign, separated by nodal lines on which the
eigenfunctions vanish.  The statistical properties of these nodal
domains then constitute a natural way to characterize the
morphology of the eigenfunctions.

Nodal domain statistics were studied for separable billiards in
\cite{BGS1}, where it was shown that if $\nu_n$ is the number of
nodal domains in the $n$th energy eigenstate then $\chi_n=\nu_n/n$
has a limiting distribution as $n\rightarrow\infty$ with a
square-root singularity at a system-dependent maximum value
$\chi_{\rm max}<1$.

In chaotic systems the eigenfunctions may be modeled
statistically, far from boundaries and turning points, by random
superpositions of plane waves \cite{Berry}:
\begin{equation}
u(\x)=\sqrt{\frac{2}{J}}\sum_{j=1}^{J}
\cos(kx\cos\theta_{j}+ky\sin\theta_{j}+\phi_{j})\label{ranwave}
\end{equation}
where $\theta_j$ and $\phi_j$ are random phases.  This is known as
the {\it random wave model}. Since plane waves are solutions of
the Schr\"odinger equation for a free particle,
$\nabla^2\psi=-k^2\psi$, the maxima of any superposition are
positive and the minima are negative. Hence the nodal domains
correspond to groups of either maxima or minima. A given pair of
adjacent maxima (minima) lie in the same nodal domain if the
saddle point between them is positive (negative). The density of
saddles in the random wave model is asymptotically twice the
density maxima or minima.  This would be exactly the case, for
example, if the maxima and minima lay on alternate sites of a
square lattice and the saddles on the corresponding dual lattice,
e.g.~midway between diagonally adjacent maxima (or minima)
(although it is important to note that in the random wave model
typical realizations of the wave function are in fact highly
irregular). The saddles may be thought of as lying at the
midpoints of {\it bonds} of the dual lattice connecting the
maxima, for example.  If the saddle height is positive, then the
corresponding maxima are connected and the bond may be thought of
as 'open'; if it is negative, the maxima are not directly
connected, and the bond may be thought of as 'closed'. This was
the basis of the very interesting suggestion put forward by
Bogomolny and Schmit \cite{BS} that statistical properties of
nodal domains in the random wave model, and hence in quantum
chaotic eigenfunctions, correspond to those in {\it critical
percolation}.  Specifically, Bogomolny and Schmit assumed that the
heights of the saddles are uncorrelated and have equal probability
of being positive or negative, and proposed {\it bond percolation
on a square lattice} as a model for nodal domain statistics. This
implies that $\chi_n$ is Gaussian distributed as $n$ varies in the
semiclassical limit. Moreover, it leads to the conclusion that the
scaling exponents associated with critical percolation also
characterize properties of the nodal domains in quantum chaotic
eigenfunctions, for example their area distribution and fractal
dimension.

The predictions of the percolation model are consistent with
numerical computations \cite{BS} and experimental measurements
\cite{micro} of the fluctuation statistics for the nodal domains
of quantum billiards, but the data do not provide conclusive
verification.  This is important, because the model has been the
subject of considerable debate.  Foltin has shown that the heights
of the saddles in the random wave model exhibit long range
correlations \cite{Foltin03}, contradicting one of the key
assumptions of the percolation model.  Bogomolny has argued that
oscillations in the two-point correlation function are sufficient
to ensure the applicability of the Harris criterion \cite{Harris}
and so guarantee that the scaling exponents are unaffected
\cite{Bogomolny}, but the issue awaits a more systematic
investigation.  Moreover, Foltin, Gnutzmann and Smilansky have
devised a particular statistical measure for which the percolation
model fails \cite{Foltetal}.  The range of validity of the model
and the precise assumptions upon which it relies thus remain to be
determined.

Our purpose here is to establish a percolation model for quantum
torus maps.  These are some of the most important examples of
quantum chaotic systems, because one can find maps that are fully
chaotic and quantum mechanically they are finite dimensional and
so easily tractable.  We will show that for these systems there is
a critical percolation model that follows directly from the
Bohigas-Giannoni-Schmit (BGS) conjecture, which asserts that local
quantum fluctuation statistics in classically chaotic systems are
modeled by Random Matrix Theory \cite{BGS}. This model corresponds
to {\it site percolation on a triangular lattice}, which falls
into the same universality class as bond percolation on a square
lattice and so has the same critical exponents. The advantages of
investigating the percolation model for maps are, first, that the
assumptions underlying it are very much more straightforward --
one only has to assume the BGS conjecture, and there are no
problems analogous to those relating to the slow decay of
correlations in billiard eigenfunctions -- and, second, that one
can perform more extensive and controlled numerical computations,
leading to significantly more precise tests of the predictions.

We find that the percolation model for maps is extremely accurate
in that the critical scaling exponents associated with the nodal
domains are very close to those predicted by percolation theory.
Moreover, the agreement goes beyond scaling laws: the nodal
domains of the quantum maps we study also obey the Cardy crossing
formula and its generalizations.  We also verify that Cardy's
formula is satisfied within the random wave model.  This suggests
that both linear superpositions of random waves and quantum
chaotic eigenfunctions may exhibit conformal invariance in the
semiclassical limit. Finally, the link between processes governed
by Stochastic Loewner Evolution (SLE) and statistical models has
recently been the focus of considerable attention. Critical
percolation is believed to relate to SLE with diffusion constant
$\kappa=6$. For percolation on a triangular lattice this has been
established rigorously \cite{Smirnov}.  On the basis of the
percolation model one would expect nodal lines to behave like
processes governed by ${\rm SLE}_6$.  We find evidence that this
is the case for the boundaries of the nodal domains in the case of
quantum maps.

The systems we study correspond to chaotic sympletic maps acting
on the unit 2$L$-dimensional torus, which is viewed as their phase
space.  Such maps can be quantized using an approach introduced by
Hannay and Berry \cite{HB}.  The Hilbert space has finite
dimension $N^L$, where $N$ plays the role of the inverse of
Planck's constant. Quantum maps correspond to unitary matrices $U$
acting on wave functions in this Hilbert space so as to generate
their (discrete) time evolution. In the position representation
these wave functions take values on an $L$-dimensional lattice.
For example, when $L=1$ the wave functions take values $(c_1, c_2,
\ldots, c_N)$ at positions $q=Q/N$, $0\le Q<N$; and when $L=2$
they take values $(c_1, c_2, \ldots, c_{N^2})$ at positions on the
square lattice $\mathbf{q}=(Q_1/N, Q_2/N)$, $0\le Q_1, Q_2<N$.  We
shall be concerned with the quantum map eigenvectors.  If the map
is time-reversal symmetric (and so $U$ is symmetric), the
components of the eigenvectors are real.  For a given eigenvector,
we can thus split the quantum lattice into regions such that the
components associated with neighbouring sites have the same sign.
These regions then correspond to nodal domains.

When $L=1$ this can be done straightforwardly: if sites lying next
to each other on the 1-dimensional lattice have eigenvector
components $c_j$ with the same sign then they constitute part of
the same nodal domain. When $L=2$ the situation requires more
careful consideration, because one needs a convention for deciding
whether lattice points that are diagonal neighbours and have
eigenvector components with the same sign lie in the same nodal
domain or not.  Consider, for example, when the eigenvector
components associated with a group of four lattice points which
form a square have signs in a checkerboard arrangement, e.g. on
the top row + -, and underneath - +.  Are the pluses automatically
part of the same nodal domain, or the minuses?  We take as our
convention that lattice points are connected along diagonals
running from the top left to the bottom right; so in the example
just given it is the pluses that are connected.  This takes us
from the original square lattice to the {\it triangular lattice}.
Nodal domains then correspond to regions on this triangular
lattice in which connected points have the same sign. Our
convention is, of course, one of many possibilities.  However, we
note that it is necessary to incorporate diagonal neighbours for
the definition of nodal domains to be consistent with that in
billiards, and that all of the conventions we have tested which do
this lead to the same results.

In order to develop a statistical model for the nodal domains we
now need to introduce a statistical ansatz for the signs.
According to the BGS conjecture, for generic, classically chaotic,
time-reversal-symmetric systems statistical properties of the
matrix $U$ should coincide with those of random matrices taken
from the Circular Orthogonal Ensemble of Random Matrix Theory. The
joint probability density for the eigenvector components $(c_1,
c_2, \ldots, c_{N^L})$ is then
\begin{equation}
P(c_1, c_2, \ldots,
c_{N^L})=\delta\left(\sum_{j=1}^{N^L}c_j^2-1\right)
\end{equation}
that is, the eigenvectors are uniformly distributed on the unit
hypersphere.  Crucially for us, it follows immediately that the
sign of a given component is equally likely to be positive or
negative and that these signs are independent of each other at
different sites, i.e. they are {\it uncorrelated}.

When $L=1$ this model was explored in \cite{KMM}.  When $L=2$ it
corresponds directly to {\it site percolation on a triangular
lattice}, which falls into the same universality class as the
Bogomolny-Schmit model. This means that the critical exponents
associated with the nodal domain statistics will be the same.  We
note that in our case these have been established rigorously for
percolation \cite{Smirnov}.

We now test the percolation model for a particular family of
quantum torus maps.  In essence, we are seeing whether this family
is described sufficiently accurately by RMT for the model to
apply. Linear maps are not sufficient for our purpose: because of
non-generic arithmetical symmetries they are not described by RMT
\cite{Keating91}.  Instead, we take a linear map composed with a
nonlinear perturbation.  Specifically, we use $M=\rho \circ A
\circ \rho$ with
\begin{equation}
 A : \left( \begin{array}{c} q_{1}\\q_{2}\\p_{1} \\
p_{2} \end{array} \right) \rightarrow
\left( \begin{array}{cccc} 2 & -2 & -2 & -1 \\ -2 & 6 & -1 & 0\\
16 & -39 & 2 & -2\\ -39 & 94 & -2 & 6 \end{array} \right) \left(
\begin{array}{c} q_{1}\\q_{2}\\p_{1} \\ p_{2} \end{array} \right)
\mathrm{mod} 1
\end{equation}
and $\rho$ a nonlinear periodic shear in momentum: $p_1\rightarrow
p_{1}+\frac{k_{1}}{4 \pi}\cos(2 \pi q_{1})$, $p_2\rightarrow
p_{2}+\frac{k_{2}}{4 \pi}\cos(2 \pi q_{2})$.
The map $M$ is time-reversal-symmetric and, for sufficiently small
values of the perturbation parameters, completely chaotic.   The
corresponding quantum map, a unitary matrix of dimension $N^2$ can
be written down easily using the prescriptions in \cite{HB} and
\cite{KMR} (the formula is long and so we do not give it here, see
\cite{KMW} for details).  We now compare statistical properties of
the nodal domains of this map with those of percolation clusters.

\begin{figure}
\centering
\includegraphics[width=7cm,clip=true]{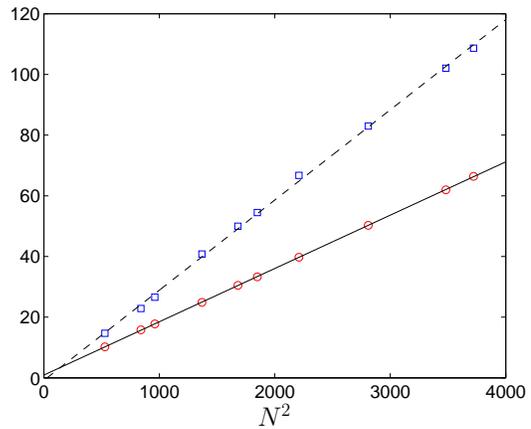}
\caption{The mean (red circles) and variance (blue squares) of the
number of nodal domains for the quantum map with $k_{1}=0.04$ and
$k_{2}=0.01$.  The linear fit for the mean (solid line) gives
$n_{c}=0.0176$ and $b=0.902$, while the fit for the variance
(dashed line) gives $c=0.0297$.} \label{fig1}
\end{figure}

\begin{figure}[h]
\centering
\includegraphics[width=7cm,clip=true]{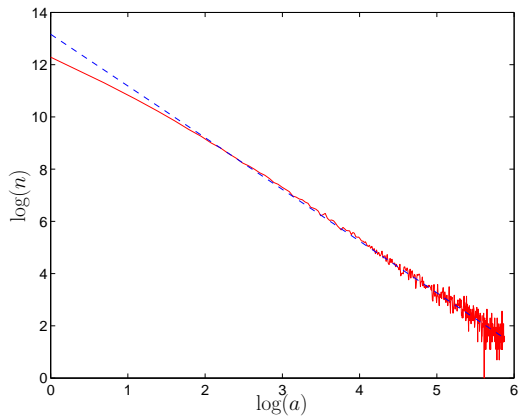}
\caption{Frequency of nodal domains as a function of area. The
solid (red) line shows the data and the dashed (blue) line the
theoretical power law with exponent $\tau=187/91$.  Here $N=61$,
$k_{1}=0.04$ and $k_{2}=0.01$.} \label{fig2}
\end{figure}

\begin{figure}
\centering
\includegraphics[width=7cm]{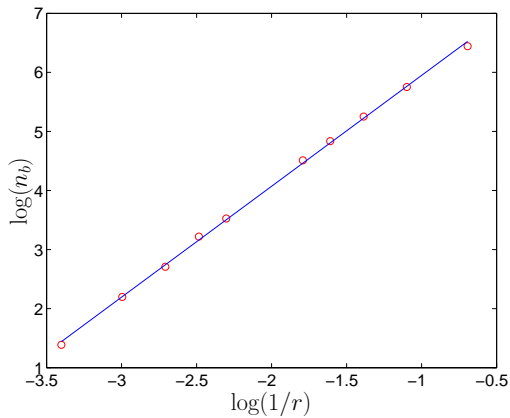}
\caption{A box count for the largest nodal domain of an
eigenvector of the map with $N=61$, $k_{1}=0.01$ and $k_{2}=0.02$.
The linear fit corresponds to a fractal dimension of $1.8774$.}
\label{fig3}
\end{figure}

Consider first the distribution of the number $n$ of nodal
domains.  For percolation on $N^2$ sites this should be a Gaussian
with mean $n_cN^2+b+o(1)$ and variance $cN^2$, where Monte Carlo
simulations give $n_c=0.0176\ldots$, $b=0.878\ldots$ and
$c=0.0309\ldots$ \cite{Ziff}.  For the quantum map we find a
Gaussian distribution with a mean and variance consistent with the
percolation formulae. Specifically, when $k_{1}=0.04$ and
$k_{2}=0.01$, a best fit gives $n_c=0.0176$, $b=0.902$ and
$c=0.0297$.  The data are shown in Figure \ref{fig1}.
For the distribution of areas $a$ of the nodal domains, the
percolation model predicts a scaling law $a^{-\tau}$, with
$\tau=187/91$. A log-log plot of the data for the eigenvectors of
the quantum map is shown in Figure \ref{fig2}.
The percolation model also implies that the nodal domains should
have a fractal dimension $D=91/48=1.89\ldots$.  Data for the
quantum map, obtained using a box-counting algorithm and shown as
a log-log plot in Figure \ref{fig3}, are consistent with this in
that the best fitting straight line has a gradient $1.8774$.

\begin{figure}
\centering
\includegraphics[width=7cm]{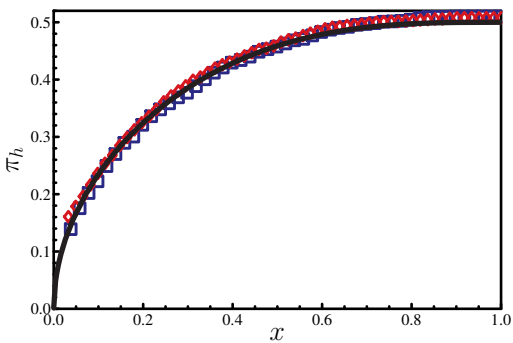}
\caption{Numerically computed crossing probabilities for map
eigenvectors when $N=61$, $k_{1}=0.01$ and $k_{2}=0.02$ (red
diamonds) and $2000$ realizations of the random wave model (blue
squares), compared with Cardy's formula (solid line).}
\label{fig4}
\end{figure}

\begin{figure}[h]
\begin{center}
\includegraphics[width=7cm]{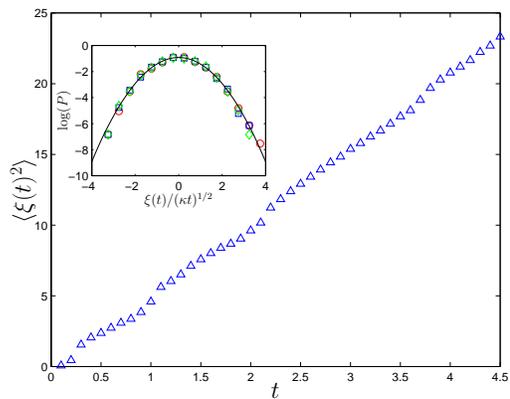}
\caption{Variance of $\xi(t)$ as a function of $t$.  The inset
shows the distribution of the rescaled driving function for
$t=0.3$ (red circles), $t=0.4$ (blue squares) and $t=0.45$ (green
diamonds) with the expected gaussian shown as a solid line.  Here
$N=61$, $k_{1}=0.02$ and $k_{2}=0.01$.} \label{fig5}
\end{center}
\end{figure}

One of the key features of critical percolation is that it has a
conformally invariant limit.  This underlies the use of conformal
field theory, in deriving the Cardy crossing formula for example,
and the link with SLE. Given the success of the percolation model
in describing scaling exponents associated with their nodal
domains, it is natural to ask whether random waves and quantum
eigenfunctions are also conformally invariant in the semiclassical
limit. In percolation, the Cardy crossing formula gives the
probability of there being a cluster spanning the system between
fixed sections of the boundary \cite{Cardy}.  For the eigenvectors
of the quantum map, we can determine the probability that a nodal
domain spans the lattice, connecting a fixed section of length $x$
on the left to any region on the right-hand side.  We also test
the nodal domains of realizations of the random wave model using
the same geometry. The results are shown in Figure \ref{fig4},
where they are compared to the crossing formula.
To explore the connection with SLE we use an idea introduced in
\cite{BGCF} to test conformal invariance in 2-D turbulence.  The
${\rm SLE}_\kappa$ process generates a trace from a stochastic
differential equation dependent on a driving function $\xi(t)$,
which is proportional to a Brownian motion.  The trace is obtained
for each eigenvector by following a nodal line while keeping the
positive points to the right. Upon hitting the left boundary the
trace continues along the boundary, in such a way that it can
always be connected to the right boundary without crossing itself,
until it reaches another nodal line. The process is stopped when
the trace reaches any of the other boundaries. The driving is
computed by approximating the candidate trace by a union of
straight line sections.  The exact solution of the Loewner
evolution for a straight line growing from the boundary \cite{KNK}
is then used to find maps which `swallow' each straight line
section in turn. This gives a sequence of times and driving
parameters. The results are shown in Figure \ref{fig5}. The
Gaussian distribution (inset) is consistent with Brownian motion.
The best fit to the diffusion constant is $\kappa\approx 5.3$.
This method is self consistent because we obtain a value of kappa
close to 6. ${\rm SLE}_6$ has the locality property that the trace
does not feel the effect of the boundary until it touches it. This
allows us to use the half-plane SLE process even in this
restricted geometry.

The fact that the crossing formula applies and the link with SLE
holds is evidence of conformal invariance.  It is natural to
conjecture that this will extend to generic quantum chaotic
eigenfunctions in the semiclassical limit.

We gratefully acknowledge stimulating discussions with Eugene
Bogomolny and Uzy Smilansky.


\begin{thebibliography}{99}
\bibitem{BGS1} G. Blum, S. Gnutzmann and U. Smilansky, {\it Phys. Rev.
Lett.}, {\bf 88} (2002), 114101.
\bibitem{Berry} M. V. Berry, {\it J. Phys. A}, {\bf 10} (1977),
2083.
\bibitem{BS} E. Bogomolny and C. Schmit, {\it Phys. Rev. Lett.}, {\bf
88} (2002), 114102.
\bibitem{micro} N. Savytskyy, O. Hul and L. Sirko, {\it Phys. Rev.
E}, {\bf 70} (2004), 056209.
\bibitem{Foltin03} G. Foltin, {\it J. Phys. A}, {\bf 36} (2003),
4561 2083.
\bibitem{Harris} A. B. Harris, {\it J. Phys. C}, {\bf 7} (1974),
1671.
\bibitem{Bogomolny} E. B. Bogomolny, unpublished.
\bibitem{Foltetal} G. Foltin, S. Gnutzmann and U. Smilansky, {\it J. Phys. A}, {\bf 37} (2004), 11363.
\bibitem{BGS} O. Bohigas, M.-J. Giannoni and C. Schmit, {\it Phys. Rev. Lett.}, {\bf 52} (1984), 1.
\bibitem{Smirnov} S. Smirnov, {\it C. R. Math. Acad. Sci. Paris}, {333} (2001),
239.
\bibitem{HB} J. H. Hannay and M. V. Berry, {\it Physica D}, {\bf
1} (1980), 267.
\bibitem{KMM} J. P. Keating, F. Mezzadri and A. G. Monastra, {\it J. Phys. A}, {\bf 36} (2003),
L53.
\bibitem{Keating91} J. P. Keating, {\it Nonlinearity}, {\bf 4} (1991),
309.
\bibitem{KMR} J. P. Keating, F. Mezzadri and J. M. Robbins, {\it Nonlinearity}, {\bf 12} (1999),
579.
\bibitem{KMW} J. P. Keating, J. Marklof and I. G. Williams, in
preparation.
\bibitem{Ziff} R. M. Ziff, S. R. Finch and V. S. Adamchik, {\it Phys. Rev. Lett.}, {\bf
79} (1997), 3447.
\bibitem{Cardy} J. Cardy, {\it J. Phys. A}, {\bf 25} (1992),
L201.
\bibitem{BGCF} D. Bernard et al., {\it Nature Physics}, {\bf 2}
(2006), 124.
\bibitem{KNK} W. Kager, B. Nienhuis and L. P. Kadanoff,
{\it J. Stat. Phys.}, {\bf 115} (2004), 805.

\end{thebibliography}
\end{document}